\pdfoutput=1
\documentclass[10pt]{article}
\usepackage[a4paper, total={6in, 8in}]{geometry}

\usepackage{xcolor}
\definecolor{purple}{HTML}{B8008A}

\newcommand{\edits}[1]{\color{black} #1 \color{black}}

\usepackage{graphicx, array, blindtext}
\usepackage{subfig}
\usepackage[export]{adjustbox}
\usepackage{enumitem}
\usepackage{url}

\usepackage{gensymb}
\usepackage{mathtools}
\usepackage{amsthm}

\usepackage{longtable}
\usepackage{authblk}

\usepackage[numbers,sort&compress]{natbib}

\begin{document}

\setcounter{secnumdepth}{0}

\title{Practical insights to thin film dewetting }

\author[1]{Karim Gadelrab~\footnote{Corresponding author, karim.gadelrab@us.bosch.com}}
\affil[1]{ Research and Technology Center, Robert Bosch LLC, Watertown, MA 02472, USA}
\author[2]{Stefan Reimann-Zitz}
\affil[2]{Research Center Pharamceutical Engineering GmbH, Inffeldgasse 13, Graz, 8010, AT}

\date{}     
\setcounter{Maxaffil}{0}
\renewcommand\Affilfont{\itshape\small}

\maketitle

\begin{abstract}

Thin liquid films exhibit rich instability and rupture dynamics that critically impact coating performance across many applications. In this work, we use lattice Boltzmann method (LBM) simulations within a lubrication-theory framework to systematically quantify how film thickness, surface energy, wettability, and intermolecular forces govern dewetting kinetics and long-time morphology. Master-curve scalings are identified for the time to dewet, revealing a strong power-law sensitivity to film thickness and a comparatively weak dependence on moderate variations in contact angle. Following rupture, the film reaches a physically meaningful coverage plateau whose magnitude correlates with material parameters and provides a practical window for morphological stabilization prior to coarsening. Long-time evolution obeys classical coarsening scaling laws, with surface energy controlling domain density. These results demonstrate that lubrication-based models can deliver predictive design guidance for evaluating coating robustness and informing materials and surface engineering strategies. 
Source code is available at https://github.com/Zitzeronion/Swalbe.jl.


\end{abstract}


\section{Introduction}

Thin liquid films are central to a variety of biological~\cite{trinschekModellingSurfactantdrivenFront2018,bruSwarmingAeruginosaLens2023, hermansLungSurfactantsDifferent2015}, physical~\cite{kondicFlowThinFilms2002,beckerComplexDewettingScenarios2003,nakamuraDynamicWettingNanoscale2013}, and technological systems~\cite{grossFluidFilmLubrication1980a,szeriFluidFilmLubrication2010,quereFluidCoatingFiber1999,dasilvasobrinhoStudyDefectsUltrathin1999,singhInkjetPrintingProcess2010,wijshoffDynamicsPiezoInkjet2010}, where their morphological stability and dewetting dynamics critically influence system behavior across scales~\cite{bonnWettingSpreading2009, oronLongscaleEvolutionThin1997, crasterDynamicsStabilityThin2009}. The thin tear fluid coating the eye can rupture (dewet) into dry spots, a critical issue in ocular biology, while biphilic surfaces on beetle shells breaks condensed thin water films into droplets and collect water from fog enabling water harvesting in arid environment. In engineering applications, thin films provide functionality in coatings, adhesives, optical and microelectronic processing, lubrication layers, and membrane technologies. Across these diverse settings, thin films exhibit a rich set of dynamical modes owing to the deformability of their free interfaces causing wave propagation, fingering instabilities, and film rupture that exposes underlying substrates~\cite{oronLongscaleEvolutionThin1997}. Thin film dewetting refers to the spontaneous break-up and retraction of a thin liquid layer from a solid substrate, leaving behind dry regions or droplets on the surface~\cite{beckerComplexDewettingScenarios2003}. At the nanoscale, destabilizing long-range van der Waals forces tend to amplify tiny thickness fluctuations, and compete with the stabilizing capillary forces to generate spinodal dewetting, self-organized hole nucleation, and highly ordered droplet patterns that underpin micro and nano fabrication routes~\cite{crasterDynamicsStabilityThin2009,allaireUsingThermalCrowding2024, kondicNanoparticleAssemblyDewetting2009}. This interplay of intermolecular interactions, capillary forces, substrate heterogeneity, and confinement is central to the modern understanding of wetting–dewetting phenomena~\cite{wasanSpreadingNanofluidsSolids2003}. Complementary high-resolution analysis of nonwetting ultrathin films further show that rupture is fundamentally controlled by molecular-scale mechanisms that give rise to singular thinning profiles and self-similar collapse dynamics independent of initial film thickness \cite{moreno2020stokes}. These studies emphasize that thin-film dewetting is not only ubiquitous, but also essential to pattern formation in nature and central to engineering applications. 

Modeling thin-film dewetting via lubrication theory leverages a long-wave asymptotic approximation to drastically simplify the Navier–Stokes equations for a slender film~\cite{reynoldsTheoryLubricationIts1886}. By assuming a large disparity between the film’s lateral extent and its thickness (small aspect ratio) and considering viscous-dominated, slow flow (low Reynolds number), the free-boundary problem is reduced to a single nonlinear evolution equation for the film thickness $h(x,t)$ that is far more tractable while still preserving the essential physics of the interface dynamics~\cite{oronLongscaleEvolutionThin1997, crasterDynamicsStabilityThin2009}. This lubrication (thin-film) model captures key physical effects driving dewetting: capillarity (surface tension) enters through curvature-induced pressure gradients that smoothen and stabilize the film, whereas long-range van der Waals forces are incorporated via a disjoining-pressure term that destabilizes ultrathin films~\cite{williamsNonlinearTheoryFilm1982, pahlavanThinFilmsPartial2018}. Lubrication-based equations have been formulated to include body forces (gravity), evaporative effects, surfactant forces, and other influences, highlighting the approach’s applicability across many thin-film settings~\cite{oronLongscaleEvolutionThin1997}. Despite its simplified assumptions, lubrication theory has shown impressive predictive capabilities in describing dewetting instabilities and film evolution~\cite{beckerComplexDewettingScenarios2003}. Linear stability analyses of the thin-film equation correctly predict the onset of rupture-producing instabilities (e.g. the fastest-growing wavelength of spinodal patterns)~\cite{crasterDynamicsStabilityThin2009}. Furthermore, nonlinear simulations using lubrication models reproduce the observed progression of dewetting: an initially uniform film develops amplified surface perturbations and eventually ruptures into dry “holes,” as the fluid withdraws into receding rims around those holes; these rims then destabilize (via a transverse Rayleigh Plateau-like instability) and break up into arrays of droplets~\cite{diezBreakupFluidRivulets2009}. Over longer times, the droplet array can undergo coarsening (smaller drops merging or fusing into larger ones), a process that the lubrication framework can also capture by reduced mean-field models \cite{kohn2002upper,otto2006coarsening}. Notably, numerous experiments have validated the lubrication-based thin-film models, finding that they successfully predict dewetting morphologies and dynamics even beyond the formal range of the theory’s strict asymptotic validity~\cite{sharmaPatternFormationUnstable1998, crasterDynamicsStabilityThin2009}. (A classic example is the spontaneous breakup of nanoscale polymer films, where the lubrication equation with van der Waals forces closely matches the pattern of holes and rims seen in experiments~\cite{beckerComplexDewettingScenarios2003,peschka2019signatures}). {A large body of work has extended the applicability of thin-film lubrication theory to regimes involving slip-dominated flows \cite{sharma1996nonlinear,munch2005lubrication,martinez2020effect}, and multiphase environments \cite{pelusi2022liquid}. More broadly, the comprehensive review by Oron et al. \cite{oron1997long} demonstrated the ability of the theory to include gravity, thermocapillarity, evaporation and condensation, surfactants, and curved or deformable substrates, while also defining regimes where long-wave reductions break down and higher-order or full hydrodynamic models become necessary.} Such theory became an indispensable tool for modeling thin-film dewetting, offering a physically insightful and computationally efficient description of how films rupture and evolve into droplets under the interplay of surface tension, intermolecular forces, and thermal effects~\cite{crasterDynamicsStabilityThin2009, bonnWettingSpreading2009, zhangMolecularSimulationThin2019}.

{In this work, we demonstrate that, despite its apparent simplicity, thin-film lubrication theory captures a remarkably rich spectrum of dynamical behavior that is directly relevant to practical coating stability and design. Building on the classical long-wave formulation, the study combines LBM simulations with systematic parametric sweeps to quantify how intermolecular forces, surface energy, wettability, and film thickness jointly govern the full dewetting pathway, from spinodal amplification and rupture to plateau formation and long-time droplet coarsening. The results establish predictive master-curve scalings for the time to dewet, identify thickness as the dominant control parameter for stability, clarify the limited benefit of moderate surface activation, and reveal a physically meaningful coverage plateau that can be exploited as a processing window for arresting morphology evolution through material or process interventions. The work illustrates how lubrication theory provides engineers with the tools to make informed design decisions for thin-film stability.}


\section{Theoretical background and implementation}


In order to model the dynamics of the dewetting liquid we use the thin film equation (TFE)~\cite{oronLongscaleEvolutionThin1997, crasterDynamicsStabilityThin2009},  
\begin{equation}\label{eq:thinsolve}
     \partial_t h(\mathbf{x},t) = \nabla\cdot\left(M_{\delta}(h)\nabla p\right),
\end{equation}
where $h(\mathbf{x},t)$ is the thickness of the film {at} $\mathbf{x} = (x,y)$ {and time $t$, and}$\nabla = (\partial_x, \partial_y)$ being horizontal components.
\edits{On the right hand side we have the mobility term $M_{\delta}(h)$ and the pressure gradient $\nabla p$, which we will discuss after the LBM scheme.}
Equation~(\ref{eq:thinsolve}) is a fourth order non-linear stiff equation that turns out to be numerical tedious. 
Instead of a direct discretization of this equation we choose a LBM~\cite{krugerLatticeBoltzmannMethod2017a, succiLatticeBoltzmannEquation2001} specifically developed for thin film dynamics~\cite{zitzLatticeBoltzmannMethod2019, zitzSwalbeJlLattice2022}.
This approach has been tested and successfully employed for, e.g., thermal fluctuating films, switchable substrates, chemically active colloids in films and ring-rivulets~\cite{ zitzLatticeBoltzmannSimulations2021, zitzControllingDewettingMorphologies2023, richterChemicallyReactiveThin2025, 10.1063/5.0256308}.
\edits{Furthermore this method has been validated against theory and other numerical tools for spreading problems, Cox-Voinov and Tanners law~\cite{cox1986dynamics, Tanner_1979}, the dewetting of a thermally fluctuating film~\cite{Grun2006, Mecke_2005,PhysRevE.100.023108, PhysRevE.92.061002} and active substrates~\cite{grawitterSteeringDropletsSubstrates2021} to name a few.}
The LBM master equation reads~\cite{chenLatticeBoltzmannMethod1998, succiLatticeBoltzmannEquation2001}
\edits{
\begin{equation}\label{eq:LBM_master}
    \begin{split}
&f_l(\mathbf{x}+\mathbf{c}^{(l)}\Delta t,t+\Delta t) = \\
&\left(1 - \frac{\Delta t}{\tau}\right) f_l(\mathbf{x},t) + \frac{\Delta t}{\tau} f_l^{(eq)}(\mathbf{x},t) + w_l \frac{\Delta t}{c_s^2} \mathbf{c}^{(l)} \cdot \mathbf{F}_{\mbox{\tiny{tot}}},
\end{split}
\end{equation}
with $f_l(\mathbf{x},t)$ are the $l$ distribution functions at lattice node $\mathbf{x}$ and time $t$, $\Delta t$ is the discrete time step, $\tau$ the relaxation time, $c_s$ is the speed of sound  and $\mathbf{F}_{\mbox{\tiny{tot}}}$ is a forcing term.
For this work we use $\Delta x = \Delta t = \tau = 1$, which sets $c_s = \sqrt{1/3}\Delta x/\Delta t = 1/\sqrt{3}$.
The lattice velocities $\mathbf{c}^{(l)}$ define the $D2Q9$ stencil we use where $l$ enumerates the nine velocities:}
\begin{equation}\label{eq:LBM_latticeCi}
    \mathbf{c}^{(l)}  =
\left\{
\begin{array}{ll}
(0,0) & l = 0 \\
\left[\cos{\frac{(l-1)\pi}{4}}, \sin{\frac{(l-1)\pi}{4}} \right] &  l=1,3,5,7 \\
\sqrt{2}\left[\cos{\frac{(l-1)\pi}{4}}, \sin{\frac{(l-1)\pi}{4}} \right] & l=2,4,6,8
\end{array}
\right..
\end{equation}

This approach essentially comprises of two steps, namely a collision step where distribution functions $f_l(\mathbf{x},t)$ relax towards their equilibria $f_l^{(eq)}(\mathbf{x},t)$ with an relaxation rate \edits{$\Delta t/\tau$ and a streaming step where}
\begin{equation}
    f_l(\mathbf{x}+\mathbf{c}^{(l)}\Delta t,t+\Delta t) = f^{\ast}_l(\mathbf{x},t),
\end{equation}
where $f^{\ast}_l(\mathbf{x},t)$ describes the post collision state.
The $w_l$ are weights and for $D2Q9$ we get
\begin{equation}
    w_l  =
\left\{
\begin{array}{ll}
\frac{4}{9} & l = 0 \\
\frac{1}{9} &  l=1,3,5,7 \\
\frac{1}{36} & l=2,4,6,8
\end{array}
\right..
\end{equation}
In contrast to the widely adopted approach to approximate the Navier-Stokes equation using a second order expansion of the Maxwell-Boltzmann distribution for the equilibria $f_l^{(eq)}(\mathbf{x},t)$ the thin film approach is build on another class of LBM solvers~\cite{salmonLatticeBoltzmannMethod1999, dellarNonhydrodynamicModesPriori2002, vanthangStudy1DLattice2010} where we have
\begin{equation}
    f_l^{(eq)}  =
\left\{
\begin{array}{ll}
h - \frac{5gh^2}{6c_s^2} - \frac{2hu^2}{3c_s^2}& l = 0 \\
\frac{gh^2}{6c_s^2} + \frac{h \mathbf{c}^{(l)}\cdot \mathbf{u}}{3 c_s^2} + \frac{h(\mathbf{c}^{(l)}\cdot \mathbf{u})^2}{2 c_s^4}-\frac{hu^2}{6c_s^2} &  l=1,3,5,7 \\
\frac{gh^2}{24c_s^2} + \frac{h \mathbf{c}^{(l)}\cdot\mathbf{u}}{12 c_s^2} + \frac{h (\mathbf{c}^{(l)}\cdot\mathbf{u})^2}{8c_s^4}-\frac{hu^2}{24c_s^2} & l=2,4,6,8
\end{array}
\right.,
\end{equation}
\edits{with $u^2 = |\mathbf{u}|^2$, $g$ being gravity (assuming that $h \sim \mu m$ or smaller we set $g \sim 0$). 
Hydrodynamic quantities such as $h$ and $\mathbf{u}$ are moments of this distribution function.}
The thickness can be computed as zeroth moment of $f_l$
\begin{equation}
    h = \sum_{l=0}^8 f_l,
\end{equation}
while the momentum is given by the first moment of $f_l$
\begin{equation}
    h u_i = \sum_{l=0}^8 c_i^{(l)}f_l,
\end{equation}
where $i = (x,y)$.
\edits{The force term $\mathbf{F}_{\mbox{\tiny{tot}}}$ in eq.~\ref{eq:LBM_master} contains the thin film mobility and the film pressure.}

The mobility $M_{\delta}$, see eq.~\ref{eq:thinsolve}, can then be computed according to,
\begin{equation}\label{eq:mobility}
    M_{\delta}(h) = \frac{1}{\mu}\left(\frac{h^3}{3} + \delta h^2 +\frac{\delta^2}{2} h\right).
\end{equation}
\edits{Here we introduce a slip length $\delta$ and use the dynamic viscosity $\mu = \rho_0\nu = \rho_0c_s^2\left(\tau-\frac{\Delta t}{2}\right) = \frac{\rho_0}{6}$.
Without loss of generality we use $\rho_0 = 1$ which in turn sets $\mu = \nu =  1/6$~\cite{zitzLatticeBoltzmannMethod2019}.}
The exact choice of this mobility is to strengthen the regularization of the contact line singularity~\cite{10.1063/5.0256308}.
In the limit of $\delta\rightarrow 0$ we recover the no-slip boundary condition, but it is also possible to recast the above expression into the Navier's slip boundary condition with an $h$-dependent effective slip length~\cite{Haley_Miksis_1991,Greenspan1978}.
The pressure (jump) at the liquid-gas interface $p$ is given by,
\begin{equation}\label{eq:filmpressure}
    p = - \gamma\nabla^2 h -\Pi(h),
\end{equation}
with a capillary contribution $\nabla^2 h$ and a so-called disjoining pressure $\Pi(h)$~\cite{schwartzSimulationDropletMotion1998, crasterDynamicsStabilityThin2009}
\edits{
\begin{equation}\label{eq:disjoinpressure}
    \Pi(h,\theta) = \frac{2\gamma}{h^{\ast}}(1-\cos\theta)\left[\left(\frac{h^*}{h}\right)^3 -\left(\frac{h^*}{h}\right)^2\right],
\end{equation}
}
with $\gamma$ being the surface tension.
{The disjoining pressure $\Pi(h)$ is the gradient of
the interfacial potential with respect to film thickness, which incorporates the interactions between liquid and substrate (i.e. wetting properties).}
Here, $h^{\ast}$ is the precursor thickness at which $\Pi(h^{\ast}, \theta) = 0$. 
Lastly, by applying the Young-Dupr\'e equation, we recover an expression for the equilibrium contact angle $\theta$ from the three surface energies~\cite{young1805iii}. 
Strictly speaking, the lubrication approximation formally requires 
$\theta \ll \pi$~\cite{oronLongscaleEvolutionThin1997, crasterDynamicsStabilityThin2009}. 
\edits{Pelusi and coauthors, however, have shown that the Navier-Stokes approximating LBM formulation with a Shan-Chen multiphase model can be used to simulate thin film dynamics~\cite{pelusi2022liquid, shanLatticeBoltzmannModel1993, shanMulticomponentLatticeBoltzmannModel1995}. 
They furthermore showed that their simulations with a disjoining pressure term generates the spectrum of a dewetting thin film, even for contact angles beyond the $\theta \ll \pi$ limit~\cite{pelusi2022liquid}.}
In this regard we explore the numerical stability of the solver for up to $\theta \approx \pi/2$.
Both eq.~\ref{eq:mobility} and eq.~\ref{eq:filmpressure} are absorbed in $\mathbf{F}_{\mbox{\tiny{tot}}}$ of the LBM approach. 

To systematically investigate the different stages of film dewetting, an extensive parametric study was conducted in which the variables appearing in eq.~(\ref{eq:disjoinpressure}) were varied over a wide range. Specifically, the dimensionless parameter ${h^{\ast}}$ was assigned values of 0.05, 0.15, 0.30, and 0.45, while the surface energy $\gamma$ was varied over the range 0.0001, 0.0005, 0.001, and 0.005. \edits{The equilibrium contact angle $\theta$ was set to $40^o$, $60{^o}$, and $80{^o}$, and the initial film thickness ${h_0}$ was taken as 2, 3, 4, and 5. All parameters are reported in LBM units, and all simulations were performed in two dimensions.}

In defining the computational domain, particular care was taken to avoid artificial periodic constraints that may arise from an insufficiently large simulation cell. \edits{To this end, a square grid of side $L = 1000\Delta x$ was employed, corresponding to approximately 7–10 times the maximum unstable wavelength ($\lambda_{max} = \sqrt{8\pi^2\gamma/\Pi'(h)|_{h_0}}$).} Random perturbations of amplitude $\epsilon/h_0  \sim O(10^{-2})$ are applied around the value of $h_0$ ($\epsilon$ being a random variable uniformly distributed in $[-1, +1]$).This setup spans a broad range of wetting and stability conditions, enabling the identification of robust scaling relations and trends relevant to a wide range of engineering and materials applications.
\edits{For the remainder of this study we normalize $h_0$ with $\Delta x$.}




\section{Results and Discussion}


Building on the theoretical framework outlined in the previous section, we now examine the numerical results characterizing the temporal evolution of thin-film dewetting. The analysis focuses on the progression from early-stage instability to film rupture and subsequent coarsening, with particular emphasis on quantifiable metrics that capture the transition between these regimes and their dependence on the governing system parameters.

\begin{figure}
    \centering
    \includegraphics[height=0.55\textwidth]{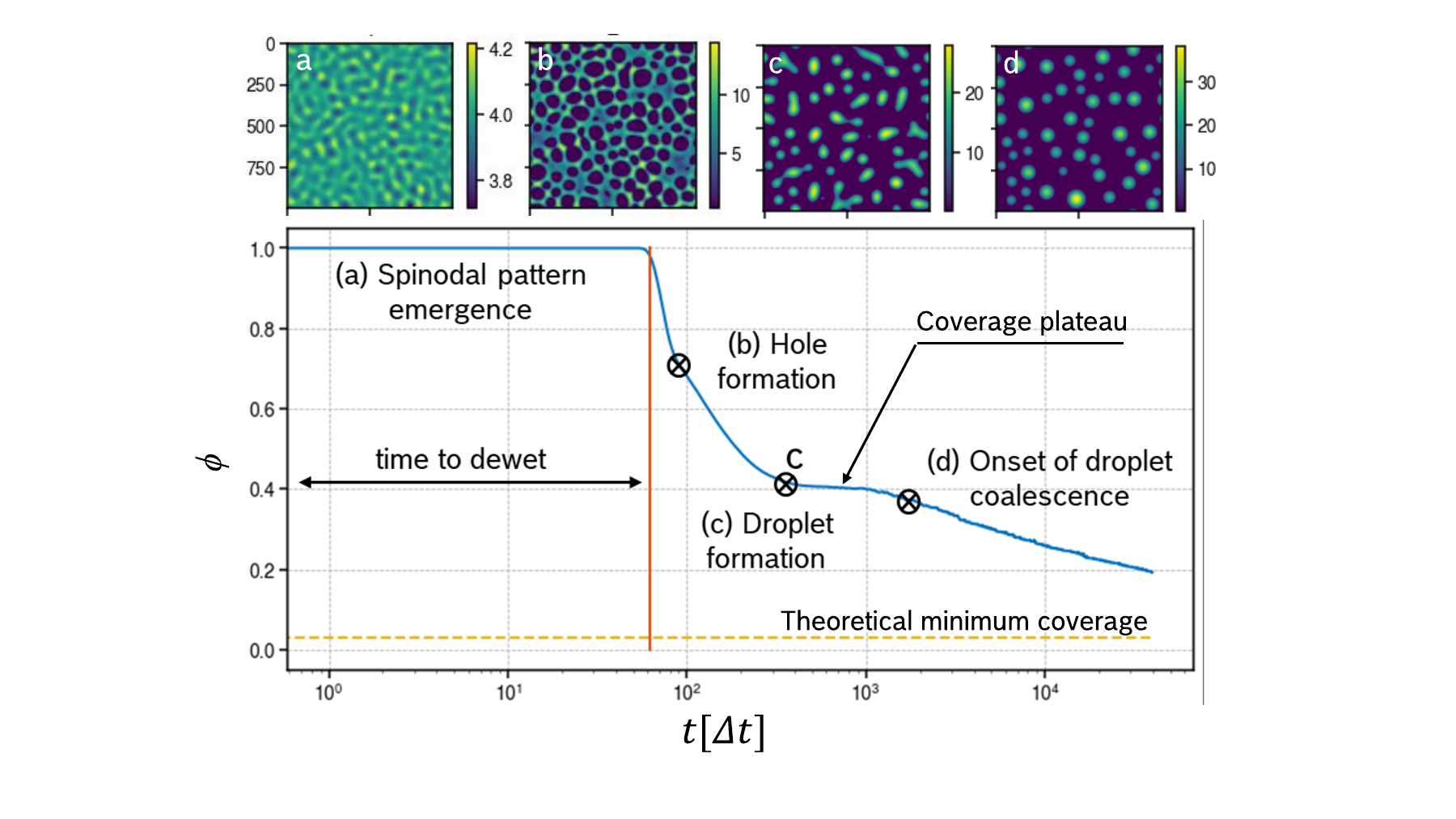}
    
    \caption{Temporal evolution of the normalized film coverage $\phi$ during the dewetting process (simulation time $t$ in LBM units). Different stages of dewetting are shown in labeled insets (a-d) and marked on the film coverage graph \edits{($\theta$ = 80$^o$, $h^*$ = 0.45, and $\gamma$ = 0.005)}. A sharp drop in film coverage indicates hole formation and marks time to dewet. Film coverage reaches a plateau where the interconnected network structure evolves to discrete droplets. A slow reduction in film coverage follows where droplets coarsening and ripening take place. Theoretical minimum film coverage $\phi_{min}$ is marked in \edits{"dashed line"}, representing the equivalent spherical cap of a droplet base area for the same film volume.
    }

    \label{fig:coverage}
\end{figure}

Fig.~\ref{fig:coverage} illustrates the temporal evolution of the normalized film coverage during the dewetting process \edits{(simulation shown for $\theta$ = 80$^o$, $h^*$ = 0.45, and $\gamma$ = 0.005)}. Initially, the film maintains complete coverage ($\phi = 1$), corresponding to a uniform state perturbed only by small-amplitude thickness fluctuations. During this stage, the system is in the spinodal regime, where long-wavelength instabilities, driven by attractive van der Waals forces and resisted by surface energy, amplify gradually while maintaining the overall continuity of the film (inset a). As these undulations grow in amplitude and approach the order of the mean film thickness, spontaneous nucleation of holes are created at the local minima in the film profile. This transition is manifested by a sharp drop in film coverage (inset b), highlighting the onset of rupture. Subsequently, the evolving holes expand and coalesce at their junctions, fragmenting the continuous film into interconnected liquid domains separated by thin capillary necks. The morphology evolves toward a configuration of discrete droplets linked by residual filaments (inset c), representing the characteristic outcome of spinodal decomposition in thin films. This is demonstrated as a continuous reduction of film coverage, albeit at a slower rate in the (b-c region). As coalescence progresses and the film reorganizes to minimize interfacial energy, the coverage reaches a secondary plateau (c-d region), marking the transition to a slow coarsening regime dominated by droplet coalescence and ripening processes (inset d). During the dewetting process $\phi$ continues to decrease asymptotically approaching the limit of an equivalent single droplet, which can be approximated as a spherical cap having the same film volume $V$ ( $\phi_{min}$= $a^2/L^2 $ where $ a^3 = 3Vsin\theta/(\pi(1-cos\theta)^2(2+cos\theta))$ ).

The time to dewetting, defined in this work as the moment when the step change in $\phi$ crosses 0.98, is expected to depend on the wetting characteristics of the film. Key factors include $\gamma$, $\theta$, and $h$, as well as the specific form of the disjoining pressure. Due to the high dimensionality of this parameter space, it is insightful to identify master curves that collapse the influence of these variables into a simple analytical relation. Establishing such correlations is essential to map the dewetting time to the system design parameters in a predictive manner. 

\begin{figure}
    \centering
    \includegraphics[height=0.55\textwidth]{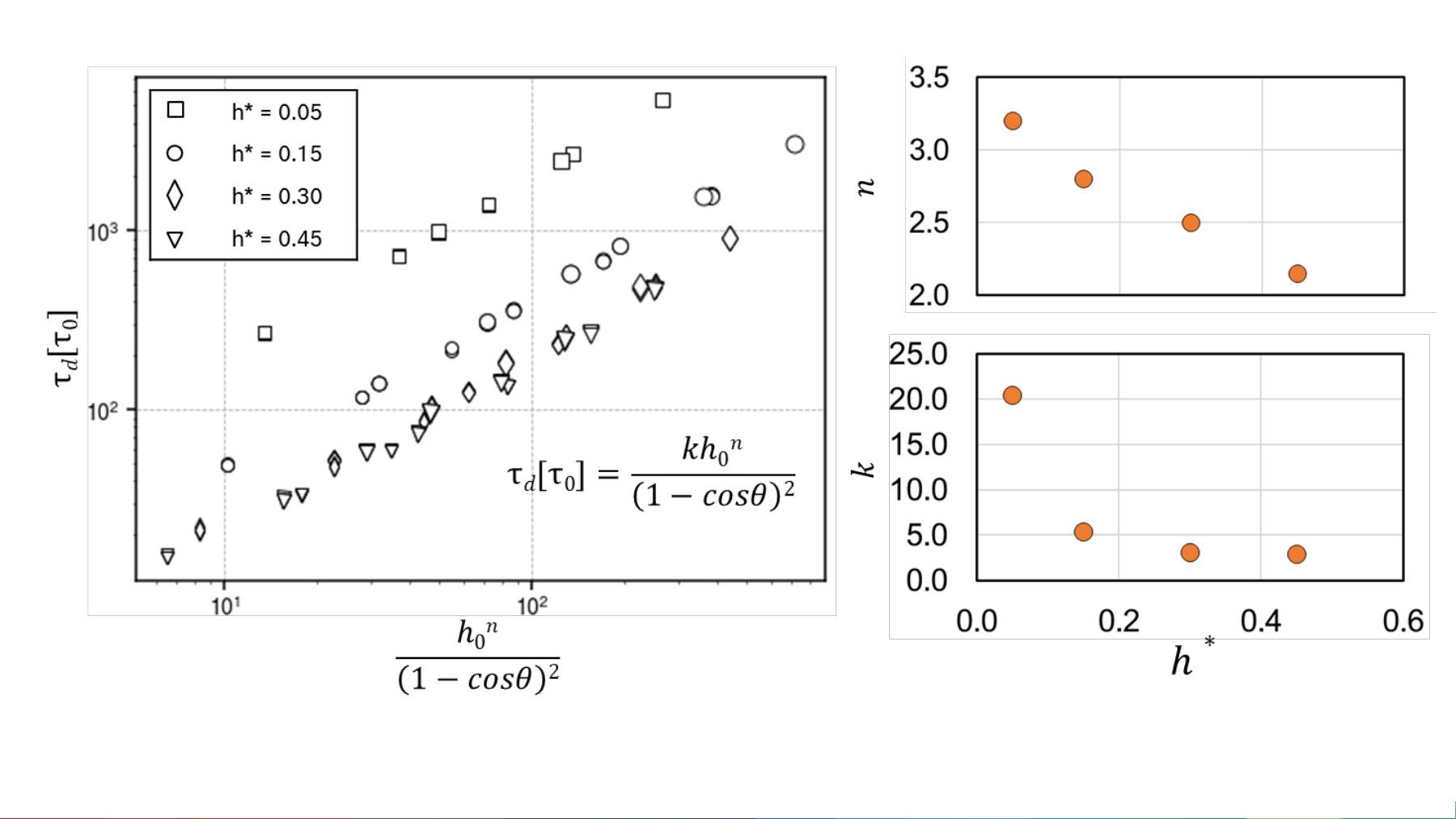}
    
    \caption{Time taken by a film to dewet as a function of the film characteristics. Universal behavior is obtained where simulation data align on separate bands that are only dependent on $h^*$. The time to dewet drops with the increase of $h^*$. A limiting behavior seems to appear for $h^*$ $\geq$ 0.3. A power law relation is obtained where the values $k$ and $n$ only depend on $h^*$ and can potentially be experimentally calibrated , or using atomistic tools such as molecular dynamics.   }

    \label{fig:scaling}
\end{figure}

Fig.~\ref{fig:scaling} demonstrates one approach to building such master curves. In this representation, the dewetting time is expressed in the form 

\begin{equation} \label{eq:masterEq}
    \tau{_d}/\tau_0= \frac{k{h_0}^n}{(1-\cos\theta)^2} 
\end{equation}
where $\tau_0 = \mu h_0/\gamma$ \edits{and $k$ is a yet unknown proportionally factor}. When $h^*$ is held constant, the data collapse onto a linear plot, indicating that the proposed scaling effectively captures the dominant parametric dependence. It is shown that the low values of $h^*$ result in a longer time of dewetting. Increasing the magnitude of $h^*$ causes the master curve to shift down (shortening dewetting time). A limiting behavior seems to appear for $h^*$ $\geq$ 0.3. The change of the proportionality constant $k$ \edits{(evident by the vertical shift of markers on a log-log plot)} for the different values of $h^*$ is plotted in Fig.~\ref{fig:scaling}, where $k$ exponentially drops and approaches a value of 2 in the limit of large $h^*$. Similarly, the exponent $n$ exhibits a monotonic decay with increasing $h^*$ starting close to a value of 3.5 for a small $h^*$ and reaches a value of 2.15 at $h^*=$ 0.45. It is worth emphasizing that the value of $h^*$ is intrinsically linked to the physical form of the disjoining pressure and, in principle, can be calibrated for specific material systems through experimental characterization or atomistic simulation methods such as molecular dynamics~\cite{diezMetallicthinfilmInstabilitySpatially2016}.

\begin{figure}

    \centering
    \includegraphics[height=0.55\textwidth]{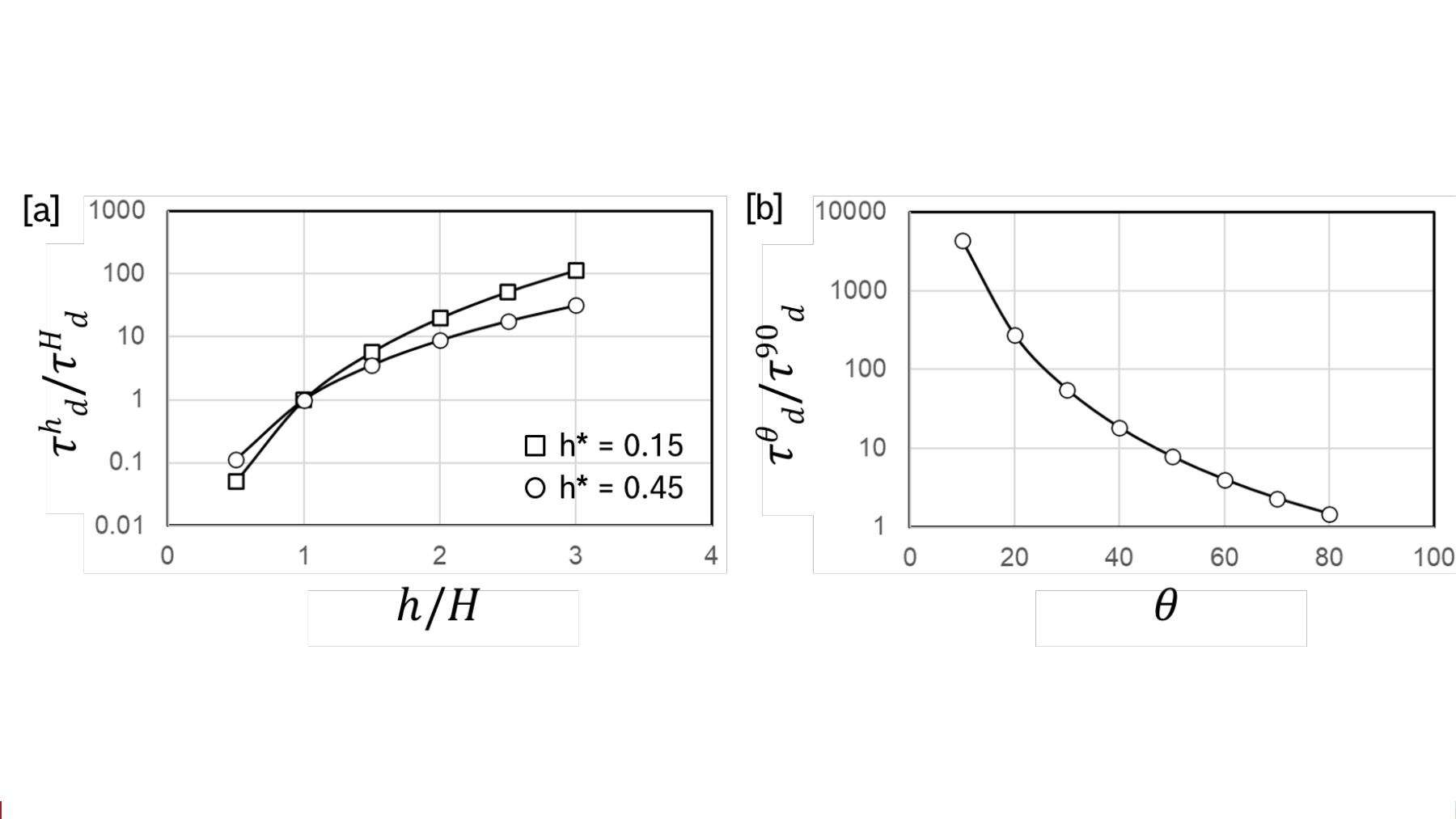}
    
    \caption{Sensitivity of $\tau_d$ to different design and fabrication changes. (a) The normalized change of $\tau_d$ due to change in film thickness ($H$ \edits{denotes the nominal (intended) film thickness set by the fabrication process}). Variations due to the lack of control on final film thickness during fabrication can result in orders of magnitude change in time to dewet. Film thickness needs to be carefully considered during the design process of coatings due to its significant impact on stability. (b) The normalized change of $\tau_d$ due to the change in $\theta$. Even for moderately compatible surfaces with the coating ($>40^o$), no significant gain in $\tau_d$ is achieved. The stability improves by orders of magnitude only when $\theta$ moves $\leq 30^o$. The plot emphasizes the fact that strategies resulting in moderate levels in surface activations and moderate reduction in $\theta$ might not be the most effective in achieving prolonged film stability.      }

    \label{fig:sensitivity}
\end{figure}

{The observed power-law dependence of the dewetting time reveals a pronounced sensitivity to $h_0$, with important implications for nanoscale systems. In such applications, precise control over film thickness is inherently difficult, and substantial spatial variations are common, particularly in the 10 nm regime \cite{orfanidi2018ink,black2017thickness,mate2013spreading}. As a result, even modest thickness variations can lead to dramatic changes in dewetting kinetics; notably, a twofold increase in film thickness would result in approximately an order-of-magnitude increase in the dewetting time (as shown in Fig.~\ref{fig:sensitivity}a)). In contrast, Fig.~\ref{fig:sensitivity}b) plots the dependence of the dewetting time on the equilibrium $\theta$. It emphasizes that substantial gains in film stability are achieved only when the surface approaches a strongly wetting  regime ($\theta$  $\leq$ 30$^o$). This behavior provides clear engineering guidance for stabilizing thin films, suggesting that impactful strategies should prioritize either precise thickness control or aggressive surface modification, through surface treatments or formulation changes, rather than incremental adjustments to moderately wetting surfaces. It is important to note that these findings are based on homogeneous nucleation, in which film rupture emerges from intrinsic thickness fluctuations across an otherwise uniform film. As such, the predicted dewetting times may be interpreted as an upper bound, since realistic systems often exhibit heterogeneous nucleation driven by defects, contaminants, surface roughness, or local chemical heterogeneities that can significantly accelerate hole formation~\cite{beckerComplexDewettingScenarios2003, thieleModellingThinfilmDewetting2003, schwartzDewettingPatternsDrying2001}. Nevertheless, the scaling relations and design guidelines identified here remain broadly applicable, as they govern the underlying hydrodynamic and capillary-driven evolution once rupture is initiated. Consequently, while absolute dewetting times can be reduced in practical applications, relative trends with respect to film thickness, surface energy, and wettability are expected to persist.}

\begin{figure}
    \centering
    \includegraphics[height=0.55\textwidth]{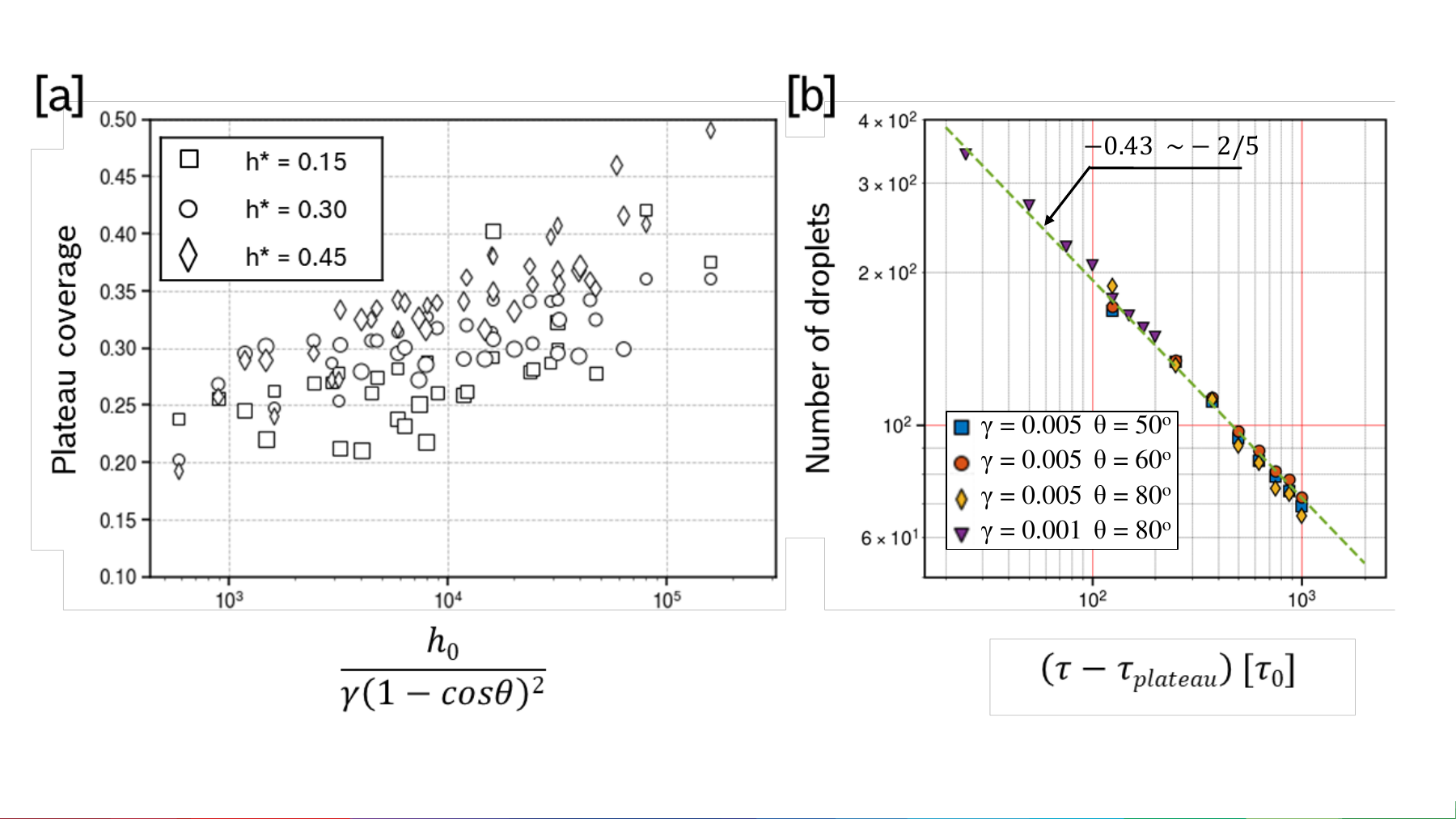}
    
    \caption{a) Quantifying plateau coverage (as marked in Fig.~\ref{fig:coverage}) at different film wetting characteristics. Plateau coverage increases proportionally with $h_0$, suggesting that thicker films retain a larger fraction of covergae after dewetting. Coverage exhibits an inverse relationship with both $\gamma$ and $\theta$, consistent with the formation of large number of beads in highly non-wetting systems. Bands based on $h^*$ are shown as systems with higher $h^*$ values tend to stabilize at higher coverage levels.
    b) Number of fluid domains (droplets) evolution with time. Coarsening follows a $\tau^{-0.43}$ scaling, which is close to the theoretical value of $\tau^{-2/5}$. \edits{The evolution of} number of domains depends weakly on $\theta$, while $\gamma$ plays a dominant role (data shown for $h_0 = 2$ and $h^* = 0.45$).}

    \label{fig:coarsening}
\end{figure}

We now extend the discussion to the long-time evolution of the film thickness and coverage after rupture. Following the initial phase of spinodal amplification and hole growth, the system reaches a quasi-steady state where the coverage stabilizes to a plateau. This plateau represents the fraction of the substrate that remains covered by the liquid domains once discrete droplets and connecting capillary filaments have formed. Interestingly, this steady-state coverage shows a measurable dependence on the governing design parameters ($h_0, \gamma,\theta $). Quantitatively (see Fig.~\ref{fig:coarsening}a), the results indicate that the plateau coverage increases proportionally with $h_0$, suggesting that thicker films, take longer to dewet, but also retain a larger fraction of covergae after dewetting. Conversely, the coverage exhibits an inverse relationship with both $\gamma$ and $\theta$, consistent with the formation of large number of beads in highly non-wetting systems. Despite some data scatter, the trends reveal clear bands based on $h^*$ ; systems with higher $h^*$ values tend to stabilize at higher coverage levels. This correlation highlights the role of $h^*$ as a unifying descriptor that encapsulates the interplay between intermolecular forces and film geometry, and it reinforces that the late-stage morphology, even after extensive rupture and coalescence, retains a strong memory of the initial wetting conditions.

{Practically, the existence of a plateau in film coverage has important implications for both characterization and control of thin films dewetting . As shown in Fig.~\ref{fig:coarsening}a, films consistently stabilize at intermediate coverages $<$ 50\%, depending on $h_0$ and wetting characteristics. This transiently arrested state provides a physically meaningful snapshot of the post-rupture morphology, reflecting the balance between capillary-driven retraction and volume conservation before long-time coarsening dominates. This regime can be exploited as a design window in practical systems. By deliberately modifying material or processing parameters, such as increasing viscosity via additives \cite{martini2018review}, accelerating solvent removal \cite{jin2017nanopatterning,bai2015situ}, or introducing chemical stabilization mechanisms including crosslinking \cite{tani2012bonding, zoppe2017surface}, it may be possible to arrest the film evolution at or near the plateau state. Such strategies could effectively suppress subsequent droplet coarsening and morphological degradation, enabling enhanced control over the final film structure in applications where partial coverage or stable droplet arrays are desired.}
 
The long-time behavior of dewetting films is dominated by coarsening dynamics, in which the number of droplets $N(\tau)$ decreases following a scaling law of $N(\tau) \sim \tau^{-2/5}$~\cite{pahlavanThinFilmsPartial2018, glasnerCollisionCollapseDroplets2005}. This exponent originates from mass-conserving coalescence events where smaller droplets collapse and feed larger ones through thin-film mass fluxes. As the system evolves, the total number of droplets decreases while their characteristic size and separation grow in a self-similar manner, preserving the total film volume. These results underscore that the late-stage coarsening process is universal, largely independent of microscopic details of the initial rupture, and primarily controlled by hydrodynamic mass transport between quasi-steady droplets.

\edits{Applied to Fig.~\ref{fig:coarsening}b, which depicts the change of number of fluid domains corresponding to the late time evolution of the normalized film coverage, this theoretical framework correlates the gradual decay of coverage as a signature of coarsening. Due to the excessively long simulation time required to access a reliable coarsening regime, we limited the conditions investigated to $h_0 = 2$ and $h^* = 0.45$. After the film ruptures and distinct droplets form, the system approaches a quasi-steady plateau where coarsening proceeds slowly with a $\tau^{-0.43}$ scaling, which is close to the theoretical value of $\tau^{-2/5}$~\cite{pahlavanThinFilmsPartial2018, glasnerCollisionCollapseDroplets2005}. The simulations suggest that the evolution of the number of domains depends weakly on the contact angle, consistent with theory, since coarsening rates and morphological self-similarity are not strongly influenced by local wetting conditions once droplets are established. Instead, the surface energy of the fluid plays the dominant role. Hence, this analysis complements the findings in Fig.~\ref{fig:coarsening}a, where smaller surface energy results in an initially large  number of domains (higher plateau coverage) depicted in the observed vertical shift in Fig.~\ref{fig:coarsening}b, with no impact on the scaling law.} 

\section{Conclusion}

In this work, we demonstrated that thin-film lubrication theory, when combined with an efficient LBM implementation, captures a remarkably rich spectrum of dewetting dynamics across a broad range of material and processing parameters. Systematic parametric sweeps revealed robust master-curve scalings for the time to dewet, highlighting film thickness as the dominant control variable and clarifying the limited benefit of moderate surface activation. Beyond rupture, the emergence of a reproducible coverage plateau and the subsequent coarsening dynamics were shown to encode meaningful information about long-time morphology, domain density, and material sensitivity. These results establish a quantitative framework for translating simplified continuum models into actionable engineering guidance for coating stability, sensitivity analysis, and process optimization.

A key practical advantage of the present approach is its accessibility and computational efficiency. The lubrication-LBM framework enables rapid exploration of high-dimensional design spaces that would be prohibitively expensive using fully resolved multiphase hydrodynamics or molecular simulations. This makes it well suited for early-stage screening, uncertainty quantification, and virtual prototyping of coating formulations and surface treatments. It would be relevant to extend the framework to incorporate additional physical effects relevant to industrial applications, including evaporation, thermocapillary stresses, gravity, surfactant transport, substrate patterning, and chemically reactive films. Furthermore, parameters such as the effective disjoining pressure or slip length can, in principle, be calibrated using targeted experiments or atomistic simulations, enabling multi-scale coupling between molecular physics and continuum predictions.

\bibliographystyle{achemso}
\bibliography{fuelCell}

\clearpage

\begin{figure}
    \centering
    \includegraphics[height=0.55\textwidth]{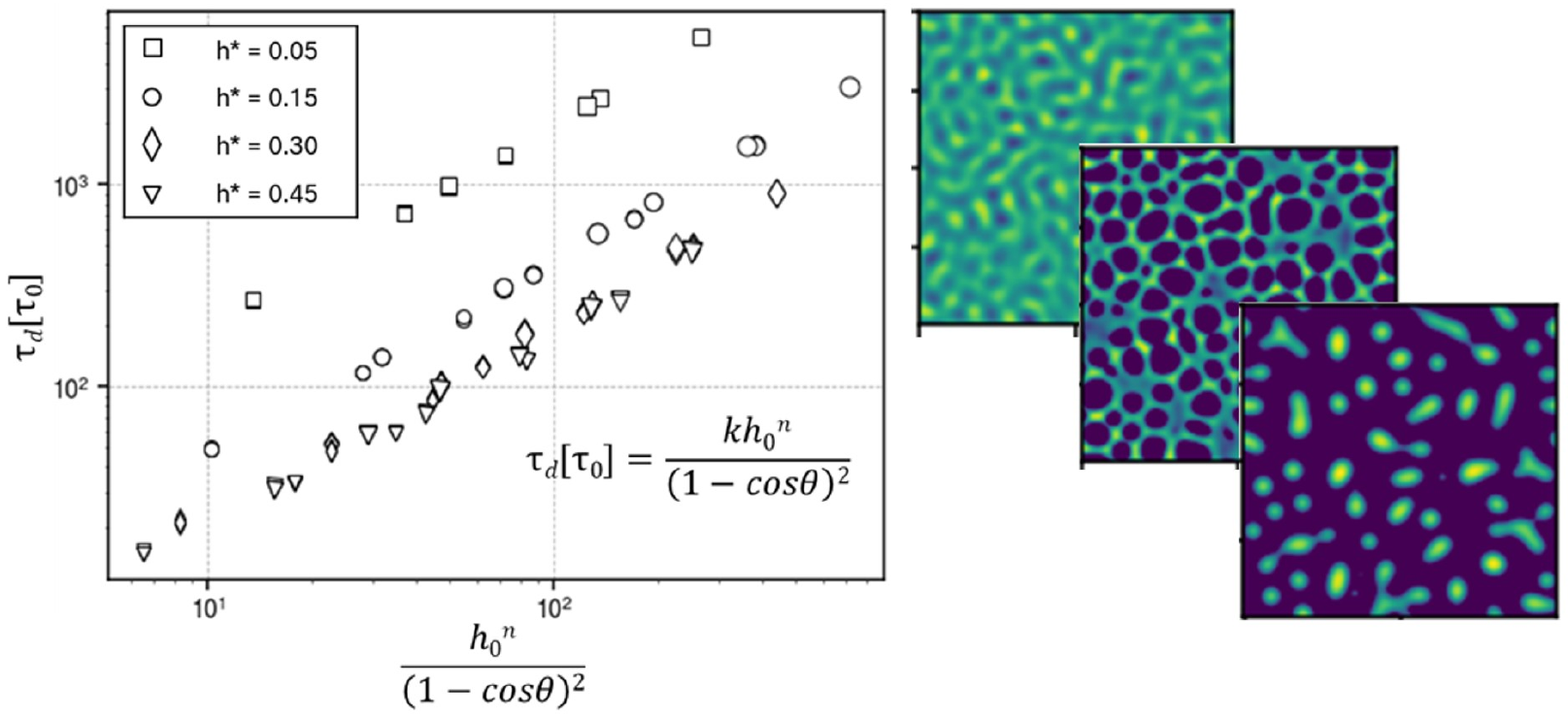}
    

    \label{fig:toc_figure}
\end{figure}

\end{document}